\documentclass{aa}  

\usepackage{graphicx}
\usepackage{txfonts}
\usepackage{xcolor}
\usepackage{tablefootnote}
\usepackage{footnote}
\usepackage{multirow}
\usepackage{hyperref}

\begin{document}

   \title{Long term monitoring of FRB~20121102 with the Nançay Radio Telescope and multi-wavelength campaigns including \textit{INTEGRAL}}

%   \subtitle{I. Overviewing the $\kappa$-mechanism}

\titlerunning{Multi-wavelength campaigns on FRB~20121102}

   \author{C.~Gouiffés\inst{1} \thanks{Deceased on February 26th, 2023}
         \and
          C.~Ng\inst{2} \thanks{Corresponding author: cherry.ng-guiheneuf@cnrs-orleans.fr}
          \and
          I.~Cognard\inst{2,3}
          \and
          M.~Dennefeld\inst{4}
          \and
          N.~Devaney\inst{5}
          \and
          V.~S.~Dhillon\inst{6,7}
          \and
          J.~Guilet\inst{1}
          \and
          P.~Laurent\inst{1}
          \and
          E.~Le Floc’h\inst{1}
          \and
          A.~J.~Maury\inst{8}
          \and
          K.~Nimmo\inst{9}
          \and
          A.~Shearer\inst{5}
          \and
          L.~G.~Spitler\inst{10}
          \and
          P.~Zarka\inst{11,3}
          \and 
          S. Corbel\inst{1}
          }

   \institute{Université Paris-Saclay, Université Paris-Cité, CEA, CNRS,  AIM, 91191, Gif-sur-Yvette, France %1
         \and
             Laboratoire de Physique et Chimie de l'Environnement et de l'Espace (LPC2E) UMR7328, Université d'Orléans, CNRS, F-45071 Orléans, France %2
         \and 
         Observatoire Radioastronomique de Nançay, Observatoire de Paris, Université PSL, CNRS, Université d'Orléans, F-18330 Nançay, France  %3
         \and 
         Institut d'Astrophysique de Paris (IAP), UMR7095, CNRS and Sorbonne Université, F-75014 Paris, France %4
         \and
         Centre for Astronomy, School of Natural Sciences, University of Galway, H91 TK33 Galway, Ireland
         \and
         Astrophysics Research Cluster, School of Mathematical and Physical Sciences, University of Sheffield, Sheffield S3 7RH, UK
         \and
         Instituto de Astrofísica de Canarias, E-38205 La Laguna, Tenerife, Spain
         \and
         AIM, CEA, CNRS, Universitè Paris-Saclay, Universitè Paris Diderot, Sorbonne Paris Citè, 91191 Gif-sur-Yvette, France
         \and
         MIT Kavli Institute for Astrophysics and Space Research, Massachusetts Institute of Technology, 77 Massachusetts Ave, Cambridge, MA 02139, USA
         \and
         Max-Planck-Institut für Radioastronomie, Auf dem Hügel 69, D-53121 Bonn, Germany
         \and
         LESIA, Observatoire de Paris, CNRS, Université PSL, Sorbonne Université, Université Paris-Cité, CNRS, Place Jules Janssen, 92190, Meudon, France
             }

   \date{Received Month ??, 2024; accepted Month ??, 2024}

% \abstract{}{}{}{}{} 
% 5 {} token are mandatory
 
  \abstract
   {The origin(s) of Fast Radio Bursts (FRBs), mysterious radio bursts coming from extragalactic distances, remains unknown. Multi-wavelength observations are arguably the only way to answer this question unambiguously.}
   {We attempt to detect hard X-ray/soft $\gamma$-ray counterparts to one of the most active FRB sources, FRB~20121102, as well as improve understanding of burst properties in radio through a long-term monitoring campaign using the Nançay Radio Telescope (NRT).}
   {Multi-wavelength campaigns involving the International Gamma-ray Astrophysics Laboratory \textit{(INTEGRAL)} satellite, the Nançay Radio Observatory, the optical telescopes at the Observatoire de Haute Provence as well as Arecibo were conducted between 2017 and 2019. In 2017, the telescopes were scheduled to observe simultaneously between September 24--29. We specifically used the Fast Response Enhanced CCDs for the optical observations to ensure a high time resolution. 
   In 2019, we changed the strategy to instead conduct ToO observations on \textit{INTEGRAL} and other available facilities upon positive detection triggers from the NRT. }
   {In the 2017 campaign, FRB~20121102 was not in its burst activity window. We obtain a 5-$\sigma$ optical flux limit of 12\,mJy\,ms using the GASP and a 3-$\sigma$ limit from OHP T120cm R-band image of $R=22.2$ mag of any potential persistent emission not associated to radio bursts. In the 2019 campaign, we have simultaneous \textit{INTEGRAL} data with 11 radio bursts from the NRT and Arecibo. % and \textcolor{red}{??} burst from Arecibo. 
   We obtain a 5-$\sigma$ upper limit of $2.7\times 10^{-7}\rm{erg\, cm^{-2}}$ in the 25--400\,keV energy range for contemporary radio and high energy bursts, and a 5-$\sigma$ upper limit of $3.8\times 10^{-11}\rm{erg\, cm^{-2}}$ for permanent emission in the 25--100\,keV energy range. 
   In addition, we report on the regular observations from NRT between 2016--2020, which accounts for 119 additional radio bursts from FRB~20121102. We present an updated fit of the periodic active window of 154$\pm{2}$ days.
   %We observe a temporal Dispersion Measure variation of \textcolor{red}{XX}\,pc\,cm$^{-3}$\,/yr and a \textcolor{red}{correlation of the observed emitting bandwidth with the fluence of the bursts.}  
   }
   {}


   \maketitle
%
%-------------------------------------------------------------------

\section{Introduction}
The Fast Radio burst (FRB) phenomenon was first discovered in 2007 by Duncan Lorimer \citep{Lorimer2007} in the re-processing of Parkes Magellanic Cloud archival data in a search targeting fast transients. FRBs are recognized by their bright, short-duration radio bursts with high dispersion measures (DMs), which point to extragalactic origins. While most FRBs appear to be one-off events, the first-ever detection of repeated bursts from FRB~20121102 fundamentally changed our understanding of FRBs, and remains one of the most-studied FRB sources in the literature.

Initially discovered by the 305-m Arecibo radio telescope PALFA survey \citep{Spitler2014}, FRB~20121102 was found to repeatedly burst  \citep{Spitler2016}. It was subsequently detected by other radio telescopes, emitting from $\sim$600\,MHz \citep{Josephy2019} to $\sim$8\,GHz \citep{Gajjar2018}.
FRB~20121102 shows a diverse range of burst morphologies \citep{Hewitt2022}. It is also one of the most active FRBs, with a burst rate as high as 122\,h$^{-1}$ detected by the FAST radio telescope \citep{Li2021}. A periodic activity window of about 157 days was reported by \citet{Cruces2021} and \citet{Rajwade2020}. 
%A updated drift rate of $-$0.147$\pm$0.014\,ms$^{-1}$ was reported by \citet{Caleb2020}, 

The repetition of FRB~20121102 enabled, for the first time, interferometric follow-up observation using the Karl G. Jansky Very Large Array (VLA). Thanks to the accurate localization capability of the VLA with a resolution better than 100\,mas, FRB~20121102 was pin-pointed to a faint dwarf galaxy with a persistent radio counterpart \citep{Chatterjee2017}.
Milliarcsecond localization capability was then achieved with the European Very Long Baseline Interferometry (VLBI) networks, where the FRB was detected to co-locate with a persistent radio source with a compact size of less than 0.7\,pc \citep{Marcote2017}.
In parallel, spectroscopic observations using the Gemini Multi-Object Spectrograph (GMOS) identified the host galaxy as a low-metallicity dwarf galaxy at a redshift of $z=0.192$ (about 3 billion light-years away), and showed that the galaxy's centre offsets the persistent radio source by 200\,mas \citep{Tendulkar2017}. 

Even though the precise nature of the progenitor of FRB~20121102 remains a mystery, these radio observations have provided important clues to help piece the puzzle together. First of all, the repeating nature of FRB~20121102 rules out any cataclysmic theoretical models. \citet{Tendulkar2017} suggested that the host galaxy of FRB~20121102 is a typical home for long gamma-ray bursts (GRBs) and superluminous supernovae (SLSNe). \citet{Bassa2017} showed from high-resolution optical imaging observations that FRB~20121102 coincides with a compact region with prominent emission lines characteristic of intense star formation. \citet{Michilli2018} observed high Faraday rotation and nearly 100\% linear polarization emission in the radio bursts, which supports the scenario where the FRB is a compact source of stellar origin embedded in extremely magnetic environments.

Similar to the case of the FRB from SGR~1935$+$2154 \citep{Mereghetti2020}, multi-wavelength detections of any counterpart or afterglow associated would provide an unambiguous answer regarding the provenance of FRBs. 
Soft X-ray observations (Chandra and XMM-Newton) taken during detected radio bursts from FRB~20121102 show no contemporaneous X-ray emission \citep{Scholz2017}.
Nonetheless, several FRB progenitor models predict extended gamma-ray emission \citep[see, e.g.][]{Murase2017}. 
The project presented here attempts to detect hard X-ray/soft $\gamma$-ray (using \textit{INTEGRAL}) and optical (using the Observatoire de Haute Provence)
%, sub-mm (using ALMA) 
counterparts to the radio emission of FRB~20121102 (using the Nançay Radio Observatory).
%\textcolor{red}{Michel D. mentioned that there is a paper that %says there is no chance to detect optical emission? Which paper is %that?}
In Section~\ref{sec:telescopes}, we describe all the telescope and instrument facilities involved in this study. In Section~\ref{sec:obs}, we present the observational set up and the data processing. We discuss the results in Section~\ref{sec:discussion}. Finally, the conclusion is provided in Section~\ref{sec:conclusion}.

%--------------------------------------------------------------------
\section{Telescope facilities} \label{sec:telescopes}

\subsection{INTEGRAL}
The International Gamma-ray Astrophysics Laboratory \textit{(INTEGRAL)} satellite was launched on October 17th, 2002. With a nominal operation lifetime of 5 years, \textit{INTEGRAL} continues to collect data to date.
\textit{INTEGRAL} has a highly elliptical (51$^{\circ}$) and large (72\,h) orbit.
There are four Science Instruments on board \textit{INTEGRAL}, including the Spectrometer on Integral (SPI), the Imager onboard Integral Satellite (IBIS), the Joint European X-ray Monitor (JEM-X) and the Optical Monitoring Camera (OMC). 
For this project, we used IBIS which makes images with a 30$\times$30\,deg$^{2}$ field of view \citep{Ubertini2003}. 
The \textit{INTEGRAL} Soft Gamma Ray Imager (ISGRI) collects photon-by-photon events in the band 15$-$1000\,keV with a time resolution around 100\,$\mu$s \citep{Lebrun2003}. 
%\textcolor{red}{(Do we use the PICsIT as well?)}
Data processed was carried out using the standard \textit{INTEGRAL Offline Scientific Analysis} (OSA) software, version 11.2 \footnote{\url{https://www.isdc.unige.ch/integral/analysis\#Software}}.

\subsection{Nançay Radio Observatory}
The Nançay Radio Observatory is located in the department of Cher in the centre of France.
Two telescopes at Nançay were involved in this project. 
\subsubsection{Nançay Radio Telescope (NRT)}
The Nançay Radio Telescope (NRT) is a Kraus-type meridian telescope, with a tiltable, planar primary reflecting surface spanning 200\,m$\times$40\,m and a secondary reflecting surface in the shape of a spherical segment spanning 300\,m$\times$35\,m. The NRT has the effective sensitivity of a 100-m class radio telescope, with a system temperature of 35\,K and a telescope gain of 1.4\,K\,Jy$^{-1}$.
The observations described in this paper were acquired using the low-frequency receiver (1.1--1.8\,GHz) at the focal plane using the FORT (Foyer Optimisé pour le Radio Télescope) receiver system, providing 512\,MHz of bandwidth centred at 1484\,MHz in 128 channels.
 These data were recorded in eight subbands, each with sixteen 4-MHz channels, using the Nançay Ultimate Pulsar Processing Instrument \citep[NUPPI;][]{Desvignes2011} with a native time resolution of 64\,$\mu$s sampled at 4 bits.
%The data is coherently dedispersed to a DM of \textcolor{red}{??}\,pc\,cm$^{-3}$ within each of the \textcolor{red}{??}-MHz channels. 
Only polarization-summed intensity data was recorded and the data was not coherently dedispersed. Offline single pulse search was conducted using the \textsc{presto}\footnote{\url{https://github.com/scottransom/presto}} software \citep{presto}.

\subsubsection{NenuFAR}
The other telescope from Nançay is the New Extension in Nançay Upgrading LOFAR (NenuFAR\footnote{\url{https://nenufar.obs-nancay.fr/en/astronomer/}}) \citep{Zarka2020}.
The core array of NenuFAR consists of 96 mini-arrays each with 19 dual polarisation antennas covering an area of about 400\,m in its longest baseline. NenuFAR can operate between 10$-$85\,MHz and it is one of the most sensitive radio telescopes below 85\,MHz. 
%NenuFAR was only involved in the 2019 campaign. 
At the time of this project, NenuFAR was still in the commissioning stage and only 56 mini-arrays were included. We recorded data using the beamformed TF (Time-Frequency) mode with the UnDySPuTeD backend \citep{Bondonneau2021}. The output dynamic spectra have not been coherently dedispersed. We kept data from 44.8--82.3\,MHz with a frequency resolution of 3.05\,kHz and a time sampling rate of 21\,ms at a bit depth of 8. Only one beam is formed at the nominal position of the FRB. The recorded data are searched offline for potential FRB signals following the procedure described in \citet{Decoene2023}.

\subsection{Observatoire de Haute Provence}
While little is known about expectations of a possible optical counterpart to an FRB, it is clear that if the duration of the optical burst is as short as the radio one, the chances to detect it with standard Charge-Coupled device (CCD) cameras are reduced \citep[e.g.][]{Yang2019}, as the signal would be diluted in the longer exposures needed to reach a significant magnitude limit. We therefore used both classical CCD cameras and Fast Response Electron Multiplying CCD (EMCCD) cameras for this campaign. 

The Galway Astronomical Stokes Polarimeter (GASP) \cite{2013ExA....36..479C,phdkyne,phdoconnor,2018IAUS..337..384O}   was mounted at the Cassegrain focus of the 1.93-m telescope of the Observatoire de Haute-Provence (OHP). 
GASP uses two EMCCD cameras with a 512$\times$512 pixel format full frame. 
For fast images with the 1.93-m telescope, 24$\times$256 pixel frames were used with a pixel scale of 0$\arcsec$.42 / pixel. Each of the EMCCD cameras was run at a nominal gain of 1000, the measured gains of the two detectors were 1330 and 1370. 
Observations were conducted in the R-band which is the most efficient setting for GASP, with a frame rate of 1102.5\,Hz, giving a frame time of 907\,$\mu$s. 

Simultaneously, we also used the classical CCD camera mounted at the Newton focus of the 1.20-m telescope. The OHP-120cm has a 2048$\times$2048 CCD, with a scale of 0"38 per pixel (used here in bin 2 mode), for a total field of view of 13.1\,arcminutes a side. 
The two-telescope set-up allows us to better discriminate any spurious signals and/or confirm any candidate. 

The first half of the night was used for calibration observations. These were taken of polarimetric standards and of the Crab pulsar, coordinated between with the 1.93-m and the 1.2-m telescopes. The Crab pulsar was observed at a frame rate of 0.9 ms. The final time series can be folded in phase with the Jodrell Bank ephemeris to create phase-resolved linear and circular polarimetric light curves.
As a 2-D polarimeter GASP also measures the background polarisation and nearby secondary polarisation standards can be measured.
The dispersion measure of FRB~20121102 implies a dispersive delay of $\approx$ 2 s between the OHP optical band and any radio burst at 1.4\,GHz. Optical transients around the expected window back extrapolated from potential radio detections can be done similarly to the procedure used in the determination of the optical enhancement associated with giant radio burst in the Crab pulsar \citep{Shearer2003}.

\subsection{Arecibo}
Observations taken with the 305 m William E. Gordon Telescope at the Arecibo Observatory were obtained as part of a Director's Discretionary Time (DDT; project code P3219).
The C-lo receiver and the PUPPI pulsar backend were used to record filterbank files between 4.1--4.9\,GHz, with a frequency resolution of 1.56\,MHz and time resolution of 10.24\,$\mu$s. The observations were coherently dedispersed to a DM of 557\,pc\,cm$^{-3}$. The \textsc{presto}-based search procedure has been described in \citet{Cruces2021}. In summary, we downsample the intensity data to 81.92\,$\mu$s and 12.48\,MHz and searched between the DM range of 507 and 606\,pc\,cm$^{-3}$ with a DM step size of 1\,pc\,cm$^{-3}$. We convolve the dedispersed time series with a template bank of boxcar filters up to 40\,ms. Candidates above a signal-to-noise threshold of 6 were inspected visually. 

%Data were recorded with the C-lo receiver and PUPPI pulsar backend. PUPPI recorded filterbank files coherently dedispersed to DM = 557 pc cm−3. The recorded bandwidth at 4.1–4.9 GHz was divided into 512 channels yielding a frequency resolution of 1.56 MHz. The time resolution of the data was 10.24 μs. 

%Before searching, the filterbank data were downsampled in time to 81.92 μs, the number of channels reduced to 64, and the total intensity (Stokes I) values are extracted. The data were searched with a simple PRESTO based pipeline (Ransom 2011), downsampled in time by a factor of 16 and dedispersed with trial DMs ranging from 507 and 606 pc cm−3 in steps of 1 pc cm−3. In order to optimize burst detection, the dedispersed time-series were convolved with a template bank of boxcar matched filters up to 49 ms. Candidate bursts were identified in the convolved, dedispersed time-series by applying an S/N threshold of 6. The resulting diagnostic plots were searched by eye, and no bursts were found.

\section{Observation campaigns} \label{sec:obs}
Two separate multi-wavelength campaigns on FRB~20121102 have been carried out. See Table~\ref{tab:log} for a summary of the observation logs.

\begin{table*}
\caption{Summary of joint $\gamma$-Ray/Optical/Radio observations on FRB~20121102.}
\label{tab:log}      
\centering                          
\begin{tabular}{c| c  c c c}        
\hline\hline                
Telescope &  Start Time & End Time  & Exposure time & Other notes\\   
 &  (UTC) & (UTC)  & (s) \\   
\hline    
\multicolumn{5}{c}{\textbf{2017 campaign}} \\
\hline
\multirow{2}{*}{\textit{INTEGRAL}} & 2017-09-24 12:22:33 & 2017-09-26 16:28:18 & 179092 & Revolution 1866\\
& 2017-09-27 06:20:13 & 2017-09-29 07:19:30 & 171675 & Revolution 1867 \\
%$\hline
%Effelsberg$^{\dagger}$\\
%Arecibo$^{\dagger}$  \\
\hline
\multirow{5}{*}{GASP@OHP} & 2017-09-25 00:26:24 & 2017-09-25 04:26:24 & 14400 & seeing=1.8" \\
& 2017-09-27 00:26:42 & 2017-09-27 04:26:42 & 14400 & seeing=1.6" \\
& 2017-09-28 00:31:25 & 2017-09-28 01:54:30 & 4985 & seeing=2.0" \\
& 2017-09-28 02:31:45 & 2017-09-28 04:31:45 & 7200 & seeing=2.1" \\
& 2017-09-28 23:59:15 & 2017-09-29 01:59:15 & 7200 & seeing=1.8"\\

\hline 
\multirow{4}{*}{OHP 120-cm} & 2017-09-24 23:28:42 & 2017-09-25 04:03:23 &  42$\times$300 & seeing=2.1" \\
& 2017-09-27 00:17:18 & 2017-09-27 02:16:40 & 23$\times$300 & seeing=2.4"  \\
& 2017-09-28 00:02:25 & 2017-09-28 03:55:20 & 42$\times$300 & seeing=2.2" \\
& 2017-09-29 00:48:05 & 2017-09-29 03:26:21 &  30$\times$300 & seeing=2.2" \\
\hline

\multicolumn{5}{c}{\textbf{2019 campaign}} \\ 
\hline
\multirow{3}{*}{\textit{INTEGRAL}} & 2019-08-30 04:07:16 & 2019-09-01 09:09:36 & 182321 & Revolution 2131 \\
& 2019-09-01 19:28:59 & 2019-09-04 00:55:56 & 184376 & Revolution 2132 \\ 
& 2019-09-04 11:16:37 & 2019-09-06 05:13:53 & 141549 & Revolution 2133 \\
& 2019-09-09 20:21:15 & 2019-09-10 23:36:51 & 98136 & Crab calibration\\
\hline

NRT$^{*}$ & 2019-08-30 06:12:57 & 2019-08-30 07:27:57 &  4500 & no burst\\ 
& 2019-08-31 06:09:01 & 2019-08-31 07:16:11 &  4031 & no burst\\ 
& 2019-09-02 06:01:08 & 2019-09-02 07:16:08 &  4499 & 1 burst\\ 
& 2019-09-03 05:57:14 & 2019-09-03 07:12:13 &  4499 & 3 bursts\\ 
& 2019-09-05 05:49:22 & 2019-09-05 07:04:19 &  4498 & 3 bursts\\ 
%& 2019-09-06 05:45:25 & 2019-09-06 07:00:24 &  4499 & 1 burst\\ 
%& 2019-09-10? \textcolor{red}{??:??:??} & 2019-09-10? \textcolor{red}{??:??:??} & \textcolor{red}{?}& \textcolor{red}{4 bursts}\\
%\hline
%Arecibo & \textcolor{red}{Ask Laura}\\
%Effelsberg$^{\dagger}$ \\
\hline
\multirow{6}{*}{NenuFAR} & 2019-09-01 06:06:00 & 2019-09-01 07:18:00 & 5380  & no burst\\
& 2019-09-02 06:02:00 & 2019-09-02 07:32:00 & 5380  & no burst\\
& 2019-09-03 05:58:00 & 2019-09-03 07:28:00 & 5380  & no burst\\ 
& 2019-09-04 05:54:00 & 2019-09-04 07:24:00 & 5380  & no burst\\
& 2019-09-05 05:52:00 & 2019-09-05 07:22:00 & 5380  & no burst\\
& 2019-09-06 05:52:00 & 2019-09-06 07:18:00 & 5140  & no burst\\
\hline
\multirow{2}{*}{Arecibo} & 2019-08-31 10:36:15 & 2019-08-31 12:36:15 & $\sim$7200 & 3 bursts \\
 & 2019-09-03 10:35:43 & 2019-09-03 12:35:43 & $\sim$7200 & 6 bursts \\
%\textcolor{red}{ALMA} & \textcolor{red}{Email sent to Anaëlle} \\
\hline
%SRT \\
%GTC/Hypercam & \textcolor{red}{Ask Vik/Andy} \\
%Magic\\
\hline
%\multicolumn{4}{l}{\footnotesize{$^{\dagger}$Data already published in \citet{Cruces2021} and are thus not further discussed in this paper.}}\\
\multicolumn{4}{l}{\footnotesize{$^{*}$Regular observations were also carried out beyond these two campaigns, see Section~\ref{sec:NRT}.}}\\
\end{tabular}
\end{table*}

\subsection{The 2017 campaign}
\begin{figure}
   \centering
   \includegraphics[width=\hsize]{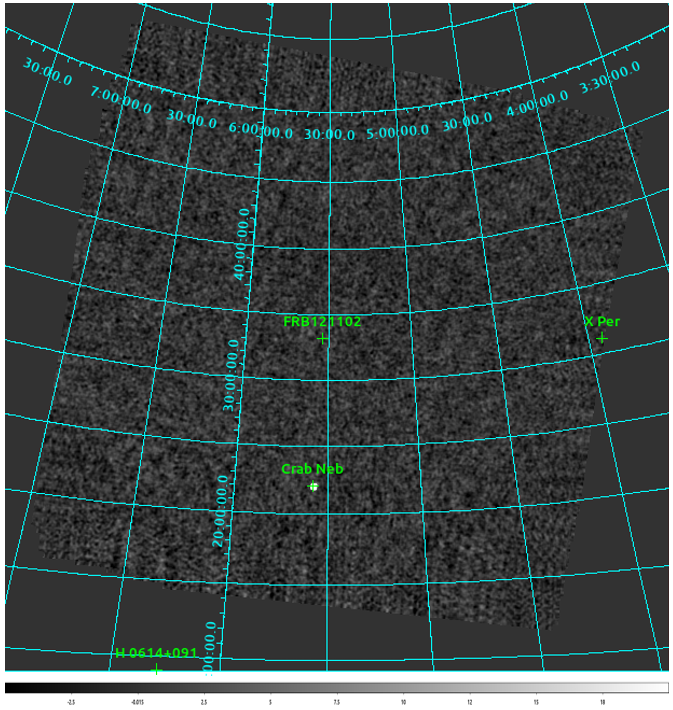}
      \caption{\textit{INTEGRAL} image at the sky position of FRB~20121102 taken during revolution 1867 at 25--100\,keV. The gray-scale bar gives the sources significance over the field of view. Only the Crab pulsar is clearly detected at more than 50 sigmas.}
         \label{fig:IntegralImage}
\end{figure}

We were granted two orbits of \textit{INTEGRAL} time on FRB~20121102 in pointing mode between September 24--29, 2017 (Proposal ID: 1420030). 
The eccentric orbit of the satellite allows continuous observation of the pointed target for nearly 3 days, except when the spacecraft trajectory enters into the radiation belts. 
An image at the sky position of FRB~20121102 taken during revolution 1867 can be seen in Fig.~\ref{fig:IntegralImage}. The upper limit of any emission that is not associated with a radio burst, at FRB~20121102 position, is $5.5\times 10^{-11}\rm{erg\, cm^{-2}}$ in the 25 -- 100 keV energy range. 

The goal of this campaign was to conduct multi-telescope observations on FRB~20121102 during this allocated \textit{INTEGRAL} period.
On the radio side, the NRT was supposed to be the primary instrument. However, due to a major failure of the hydraulic system used for the receiver motion along its track, the NRT was out of operation during this long-planned campaign. Arecibo was affected by hurricane Maria and was also not able to collect any data. 
Only Effelsberg obtained data during this campaign and this data set has been published in \citet{Cruces2021}. 
Unfortunately, no radio bursts were detected in 60\,hours of observations, which means no trigger was sent for \textit{INTEGRAL}. 
Retrospectively, this non-detection is not surprising because based on the activity window later published by \citet{Cruces2021,Rajwade2020,Li2021}, the dates of the 2017 campaign fall within a period of inactivity of FRB~20121102 (see also Fig.~\ref{fig:NRT-timeseries}).

\begin{figure}
   \centering
   \includegraphics[width=\hsize]{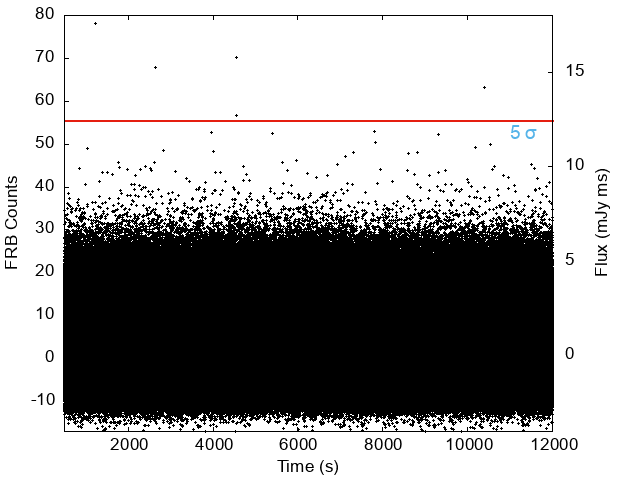}
      \caption{GASP observation of FRB~20121102 from the 2017 campaign. The plot, from the night of September 27th, shows the flux in each 0.9-ms frame. In total, there were $\approx$ 13.7 million frames. Shown are the EMCCD counts (left y axis) from the FRB region in each frame (left y-axis) and the flux in mJy ms (right y axis). Also shown is the 5 $\sigma$ upper limit. From 13.7 million frames, we expected to observe 4.1 frames with a flux above 5\,$\sigma$. The fact that we see 5 frames above 5 $\sigma$ implies no significant enhancement is detected.}
         \label{fig:GASP}
\end{figure}

\begin{figure}
   \centering
   \includegraphics[width=\hsize]{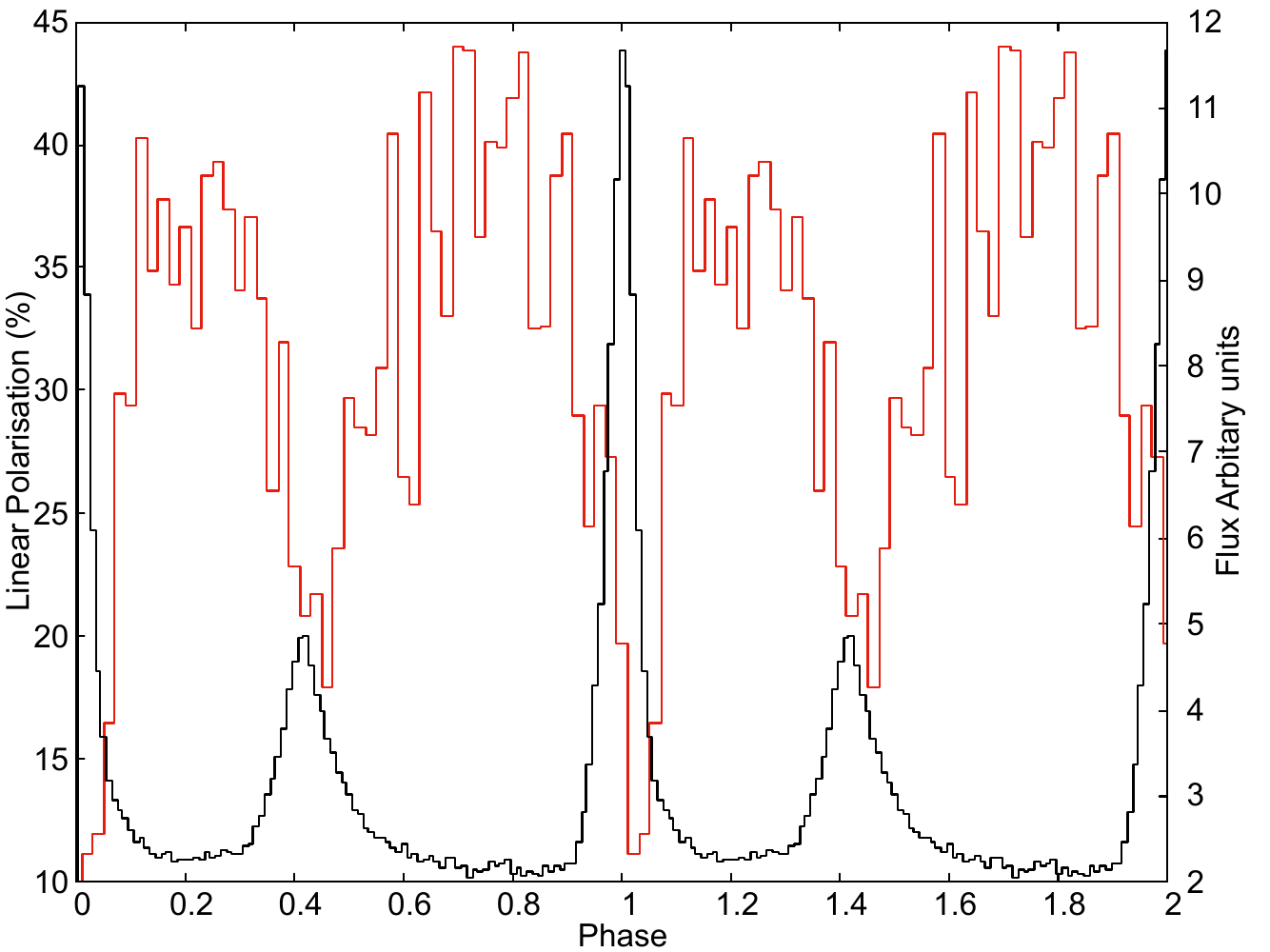}
      \caption{GASP observations of the Crab pulsar taken during the 2017 campaign. The plot shows a 1 hour observation starting at 2017:09:29 02::20:01. The red line shows the linear polarisation and the black line is the optical light curve referenced to the Jodrell Bank ephemeris, \citet{1993MNRAS.265.1003L} and 
\url{http://www.jb.man.ac.uk/~pulsar/crab.html}. The linear polarisation in the off-pulse region, from phase 0.7-0.9, is 37 $\pm$ 5 \% consistent with other measurements, see (e.g.) \citet{2009MNRAS.397..103S} }
         \label{fig:GASPPol}
\end{figure}

In the optical band, we were awarded 7 nights of OHP time, 2 of which were technical nights for the setting-up of GASP, and 2 other nights where poor sky conditions prevented observations. We were left with 4 nights (a total of 13\,h) of OHP observations of FRB~20121102 using GASP from September 25 to 29 in the 2017 campaign. 
The start and end times listed in Table~\ref{tab:log} are all UTC based on a GPS trigger with an accuracy better than 1\,$\mu$s.
Even though FRB~20121102 was in a low elevation angle, we had an 85\% efficiency in terms of weather conditions and telescope time. 
The sky was dark as the new moon was on the 19th, and the airmass was less than 1.4. 
As there were no radio events during these observations, only a 5-$\sigma$ optical flux limit could be determined at a level of 12\,mJy\,ms (Fig.~\ref{fig:GASP}). Figure~\ref{fig:GASPPol} shows the Crab light curve and provisional linear polarisation profile. 

\begin{figure}
   \centering
   \includegraphics[width=\hsize]{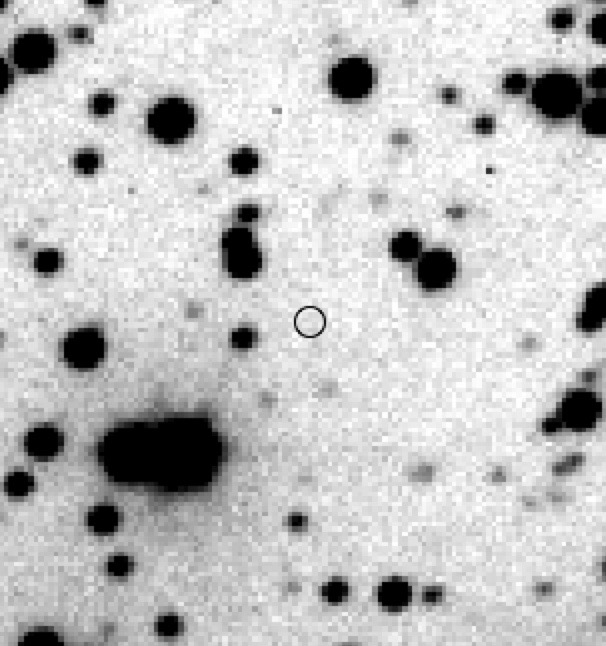}
      \caption{A 1.5\arcmin $\times$ 1.5\arcmin \, sub-section of our R-band stacked image obtained in the field of FRB~20121102 with the T120cm telescope at Observatoire de Haute-Provence. The expected location of FRB~20121102 is indicated with the empty black circle at the center. North is to the top and East is to the left. }
         \label{fig:T120}
\end{figure}

At the same time, deep R-band images of the field of FRB~20121102 were obtained with the OHP 120-cm telescope and its direct CCD camera. Consecutive exposures of 300\,s were obtained on September 24th, 26th, 27th and 28th, 2017.
Each individual image was flat-fielded and  carefully inspected but no detection was found at the location of the FRB.
The total integration time of 685\,min was used to make a stacked image.  Similarly, we do not detect any signal in the finally combined image, down to a 3-$\sigma$ limit of R\,=\,22.2\,mag (see Fig.~\ref{fig:T120}).

%% Emeric's addtion :
%%%%%%%%%%%%%%%%%%%
%%%%%%%%%%%%%%%%%%%

%Calibration observations on the Crab, blazars and white dwarf were taken in coordination with the T120cm. 

\subsection{The 2019 campaign}
We changed our approach in 2019, proposing instead to conduct Target of Opportunity (ToO) observations with \textit{INTEGRAL} (Proposal ID: 1640014). We were awarded 3 orbits each lasting about 2.5 days. Again using the NRT as the primary radio telescope, our strategy was that if radio bursts were detected in at least  2 over 3 one-hour long successive observations with the NRT, then we would have the confidence that FRB~20121102 is in an active status and \textit{INTEGRAL} would be triggered. 
Positive detections from NRT happened at the end of August, and we successfully triggered \textit{INTEGRAL} to point to FRB~20121102. In addition to the 3 orbits, calibration observations of the Crab Pulsar taken on Sept 10 also covered the location of FRB~20121102. In total, these \textit{INTEGRAL} data coincide with 11 bursts from NRT and 8 bursts from Arecibo. 
We have also sent out an Atel (ID:13073) to announce the times of the \textit{INTEGRAL} ToO observations and encouraged other facilities to take part in the follow-up observations. 

\begin{figure*}
   \centering
   \includegraphics[width=\hsize]{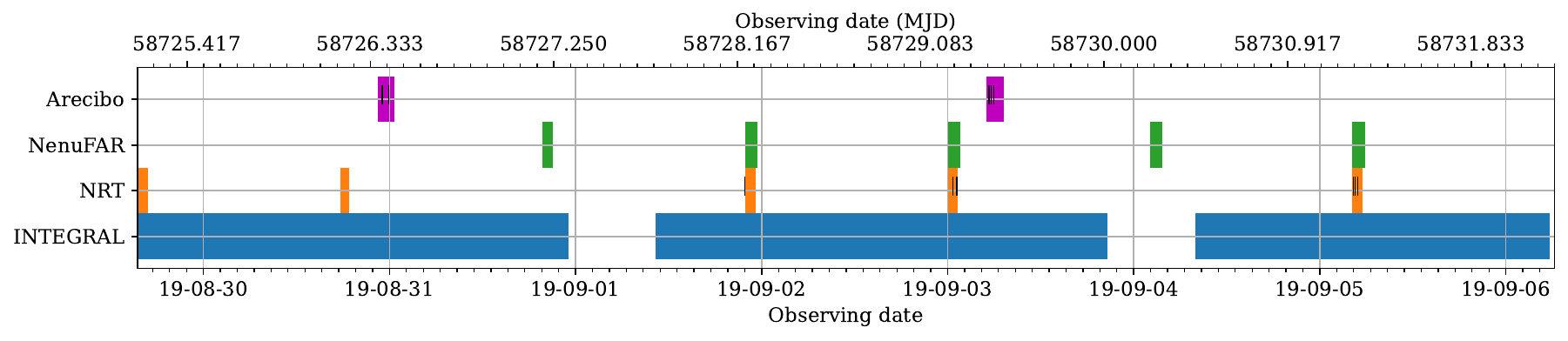}
      \caption{The time windows of the telescope facilities that participated in the 2019 campaign. Each radio burst detected is shown as a vertical black tick.}
         \label{fig:MW2019}
\end{figure*}

\begin{figure}
   \centering
   \includegraphics[width=\hsize]{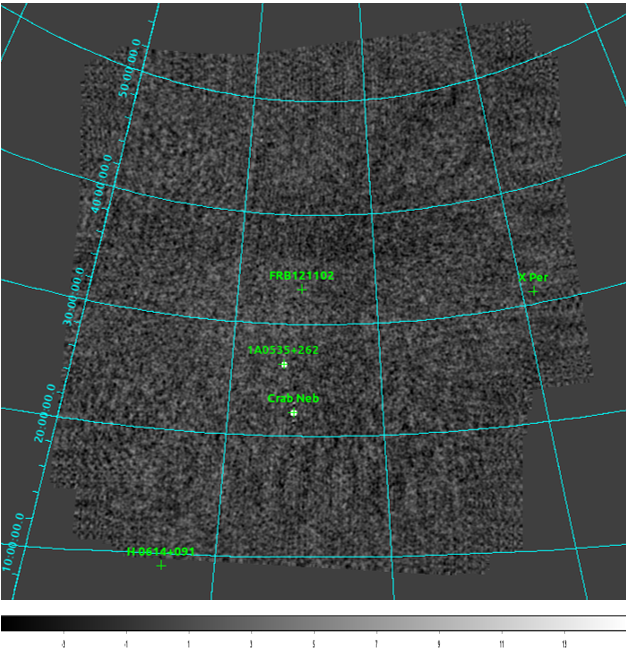}
      \caption{\textit{INTEGRAL} image at the sky position of FRB~20121102 at 25--100\,keV, stacking all observations taken during the 2019 campaign. Color bar gives the sources significance over the field of view.}
         \label{fig:IntegralImage2019}
\end{figure}

%Joint-analysis of Integral-NRT
In order to narrow down the precise burst moment in the \textit{INTEGRAL} data, we 
convert the burst time of arrivals at NRT to barycentric and 
translate the time stamps to \textit{INTEGRAL} Julian Date (IJD). 
%The fact that the Crab pulsar is within the same field-of-view meant that we had an independent way to calculate the absolute time reconstruction down to a precision of $\sim$100\,us as the phase shift of the Crab pulsar between radio and gamma-ray is well determined.
We have a total of 11 radio bursts with coincident photons during \textit{INTEGRAL} observations (see Table~\ref{tab:log}). 7 of these were during the 3 awarded \textit{INTEGRAL} orbits while 4 additional bursts came during later calibration observations of the Crab pulsar. If we consider a burst window size of 5\,ms, as it was done for the XMM Newton analysis presented in \citet{Scholz2017}, no \textit{INTEGRAL}/ISGRI emission was detected in coincidence with these 11 radio bursts, with a 5 sigmas upper limit of $2.7\times 10^{-7}\rm{erg\, cm^{-2}}$ in the 25 -- 400\,keV energy range. 
Furthermore, no permanent emission was detected from an analysis of the \textit{INTEGRAL}/ISGRI flux density during the 2019 campaign, with a 5 sigmas upper limit of $3.8\times 10^{-11}\rm{erg\, cm^{-2}}$ in the 25 -- 100\,keV energy range 
 (Refer to Fig.~\ref{fig:IntegralImage2019}).

Supporting radio observations were arranged with Arecibo, Effelsberg and NenuFAR to coincide with the NRT campaign (Fig.~\ref{fig:MW2019}). 
The NRT analysis is separately discussed in Section~\ref{sec:NRT}.
 Six consecutive days of observations were obtained at NenuFAR between September 1 to 6, 2019. On each day, a 30\,min scan was recorded on the sky position of FRB~20121102. No positive detection was made.  
A total of 9 bursts were detected from Arecibo during two $\sim$2\,hr-long observations taken on August 31 (burst MJDs: 58726.49465635, 58726.46079819, 58726.46407909) and September 3 (burst MJDs: 58729.45553347, 58729.46330519, 58729.46529836, 58729.47564011, 58729.45170017, 58729.45795891). 
Positive detections with Effelsberg were published in \citet{Cruces2021} and are not further discussed in this paper. We are aware that the FAST telescope \citep[see also, Atel ID:13064;][]{Li2021} and MeerKAT (Atel ID:13098) have also detected radio bursts from FRB~20121102 during this activity window. 
%\item Apparently we do not have XMM-Newton because the time to trigger is too short. But did we actually have an accepted proposal or did we just not try?

\subsection{NRT analysis} \label{sec:NRT}
\begin{figure*}
   \centering
   \includegraphics[width=\hsize]{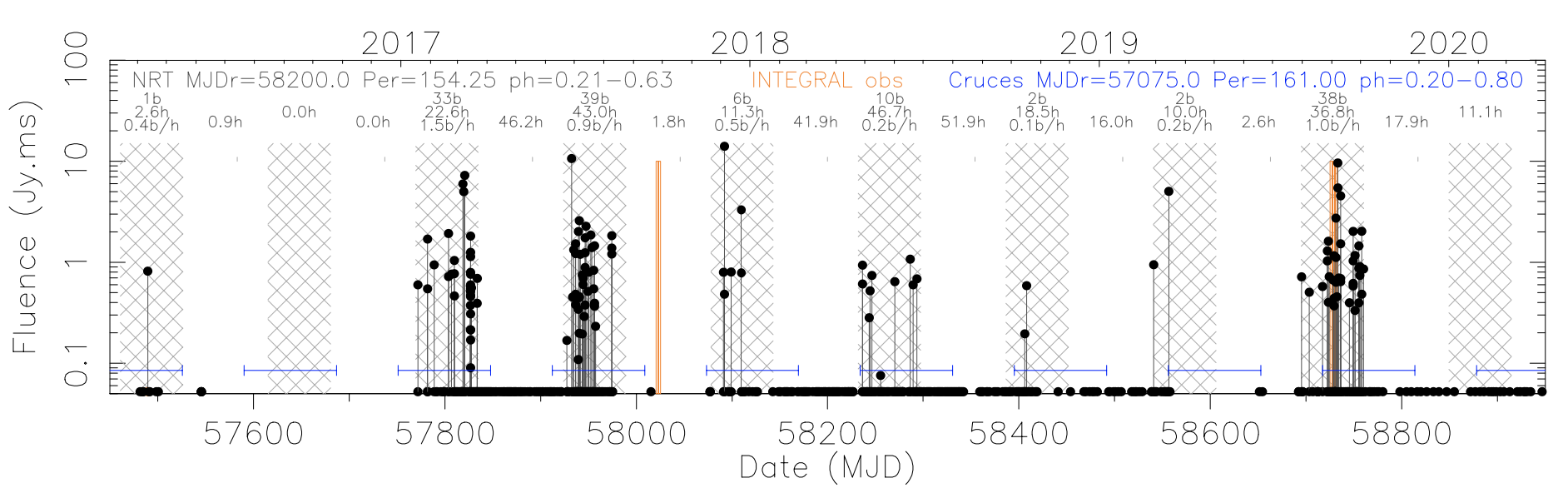}
      \caption{The fluences vs date (black dots) of the NRT detected bursts from FRB~20121102. The hatched region indicates an active window as calculated from only the NRT data, with a period of 154\,days referencing to MJD=58200. The number of bursts ($b$) detected in each active period is annotated, together with the total duration of NRT observation in each period as well as the NRT burst rate. 
      The periodicity window reported by \citet{Cruces2021} are represented by blue horizontal lines.}
         \label{fig:NRT-timeseries}
\end{figure*}

In addition to the 11 bursts simultaneous with \textit{INTEGRAL} in September 2019, a long-term monitoring campaign of FRB~20121102 took place between 2016--2020 (MJD 57481--59371) (Fig.~\ref{fig:NRT-timeseries}). This is a total of 580 observations summing up to 512.5\,h, in which a total of 131 radio bursts were detected.

\begin{table*}
\caption{Parameters of the NRT bursts. The MJDs of the burst arrival times are barycentric corrected. We list the by-eye structure-optimized DM values as well as a by-eye assessment of the emission frequency range. The measured peak burst amplitude in Jansky is also listed. }
\label{tab:NRTbursts}      
\centering                          
\begin{tabular}{cccccc}        
\hline\hline                
Burst time &  DM            & Emitting range & Emitting bandwidth & Peak Amplitude \\   
(MJD)        & (pc\,cm$^{-3}$)    & (MHz)         & (MHz)     & (Jy) \\   
\hline    
58717.301536944820  &  562.0        &  1230--1400  &  170  &  0.2\\
58722.271292859944  &  564.0       &  1230--1600  &  370  &  0.3\\
58722.294462891791  &  564.0       &  1400--1730  &  330  &  0.6\\
58723.268959840973  &  566.0      &  1250--1650  &  400  &  0.13\\
58723.305702925322  &  563.0         &  1250--1600  &  350  &  0.5\\
58724.278815864065  &  566.0       &  1300--1700  &  400  &  0.22\\
58728.251699046054  &  564.0        &  1450--1750  &  300  &  0.21\\
58729.274439107536  &  564.5       &  1280--1730  &  450  &  0.5\\
58729.293326266594  &  564.0          &  1360--1640  &  280  &  0.31\\
58729.299280584968  &  564.5       &  1420--1700  &  280  &  0.22\\
58731.250490731094  &  564.0         &  1240--1610  &  370  &  0.45\\
58731.256507079819  &  565.0          &  1240--1730  &  490  &  0.9\\
58731.271758321349  &  565.0          &  1240--1730  &  490  &  0.6\\
58732.278895409419  &  565.0          &  1240--1420  &  180  &  0.3\\
58733.243564119805  &  564.0        &  1240--1730  &  490  &  2.1\\
58733.271517283098  &  564.0         &  1240--1730  &  490  &  1.9\\
58734.252706174463  &  564.0         &  1240--1650  &  410  &  0.2\\
58736.235551187320  &  565.0        &  1240--1730  &  490  &  0.7\\
58736.240998463027  &  564.0         &  1240--1480  &  240  &  0.35\\
58736.248232336240  &  565.0         &  1340--1550  &  210  &  0.22\\
58736.254592213743  &  563.0         &  1250--1500  &  250  &  1.3\\
58745.232781696606  &  564.0        &  1320--1660  &  340  &  0.16\\
58749.198586671453  &  564.0        &  1230--1600  &  370  &  0.25\\
58749.209780319387  &  564.0          &  1300--1650  &  350  &  0.6\\
58749.213731841568  &  564.0         &  1230--1700  &  470  &  0.3\\
58749.235869732642  &  564.5       &  1250--1630  &  380  &  0.26\\
58751.199178285640  &  564.5       &  1230--1700  &  470  &  0.24\\
58751.200026219590  &  564.0       &  1230--1730  &  500  &  0.2\\
58755.190474390559  &  564.0          &  1350--1420  &  70   &  0.1\\
58755.211663606828  &  564.0         &  1300--1730  &  430  &  0.5\\
58755.213851836622  &  564.0  &  1300--1640  &  340  &  0.5\\
58756.195035061553  &  564.0         &  1230--1400  &  170  &  0.3\\
58756.223509872697  &  564.0         &  1300--1550  &  250  &  0.3\\
58758.174206763849  &  564.0           &  1230--1650  &  420  &  0.7\\
58758.221243372226  &  564.0        &  1650--1730  &  80   &  0.25\\
58760.191749209210  &  564.0           &  1450--1700  &  250  &  0.18\\
\hline 
\end{tabular}
\end{table*}

\begin{figure}
   \centering
   \includegraphics[width=2.5in,angle=270]{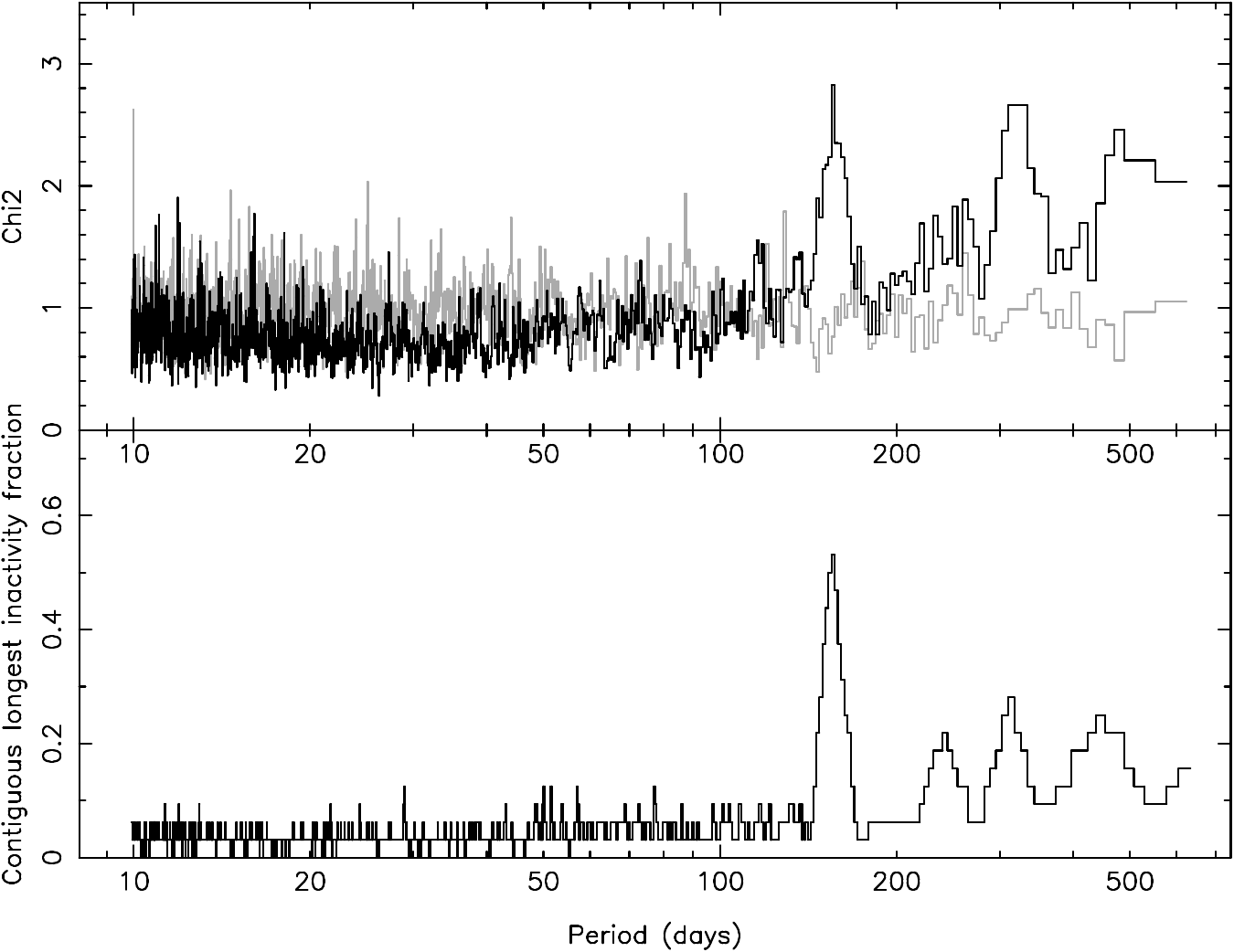}
      \caption{(top) Reduced $\chi^{2}$ (black) with respect to a uniform dis- tribution of burst arrival times for the trial periods. Reduced $\chi^{2}$ (gray) for fake burst arrival times randomly distributed among all the observing slots. (bottom) Contiguous longest inactivity fraction as a function of period trials. A clear signature at around 154 days is seen for both metrics.}
         \label{fig:Period}
\end{figure}

%Periodic window
Using the longer time span of NRT data alone, we obtain a more precise, updated periodicity window of 154$\pm{2}$ days (see Fig.~\ref{fig:Period}) with a reference MJD of 58200. The periodicity is very well constrained, as we have 131 bursts over about a 4-year span as well as over 179\,h of observations during inactive periods. 
\cite{Rajwade2020} determined a periodic activity window of 157$\pm$7 days with a reference MJD of 58200 from observations taken in 2016--2017, while \cite{Cruces2021} derived a periodic activity window of 161$\pm$5 days with a reference MJD of 57057 from 34 epochs of observations dated between 2016--2020. \cite{Li2021} finds no change to the 157-d periodicity, although that FAST data set only spans a month and a half so not very constraining on the long periodicity.
Instead, from Fig.~\ref{fig:NRT-timeseries}, we can see that at least three NRT bursts (around MJD 58550 and 58700) fall outside of the activity window predicted by previous works. 

%fitburst DM, burst width
%Temporal DM variation in FRB~20121102 has been reported by \citet{Platts2021,Hessels2019,Oostrum2020,Josephy2019}.
%The most recent analysis from the FAST team \citep{Li2021} reported DMs of 565.8$\pm$0.9\,pc\,cm$^{-3}$ between MJDs~58724 and 58776 as well as a temporal DM trend of $+$0.85$\pm$0.10\,pc\,cm$^{-3}$\,yr$^{-1}$ over a 9 year period when combining with DM data points from the literature, although the fit is highly dependent on a few early data point.
%The long time span of the NRT data set again provides an interesting update on the temporal DM variation. 
%Fig.~\ref{fig:DMtrend} shows the NRT burst DMs derived by using the \textsc{fitburst}\footnote{\url{https://chimefrb.github.io/fitburst/installation/}} code.  
%\textcolor{red}{(Work in progress)}

%\begin{figure}
 %  \centering
 %  \includegraphics[width=\hsize]{FRB121102-DMtrend.png}
  %    \caption{Temporal variations in DM vs MJD. \textcolor{red}{Waiting to add the NRT data points, the NRT pts currently on the plots are just by-eye estimates.}}
   %      \label{fig:DMtrend}
%\end{figure}

%Burst rate
The highest burst rate detected in contiguous data at the NRT is 24.4\,bursts/hr (17 bursts between MJD~57826.741875--57826.770880).
This is less than the 122\,bursts/hr observed by the FAST telescope \citep{Li2021}; not a surprise as FAST has much higher sensitivity compared to the NRT. 
The NRT and FAST burst rates are a lot higher than, for example that detected by CHIME between the lower observing frequencies of 400--800\,MHz, where only one burst has been recorded during multiple years of operation  \citep{Josephy2019}. As suggested by \citet{Houben2019}, this is possibly explained by a flattening of the spectrum of FRB~20121102 below 1\,GHz.

\begin{figure}
   \centering
   \includegraphics[angle=0,width=1.05\hsize]{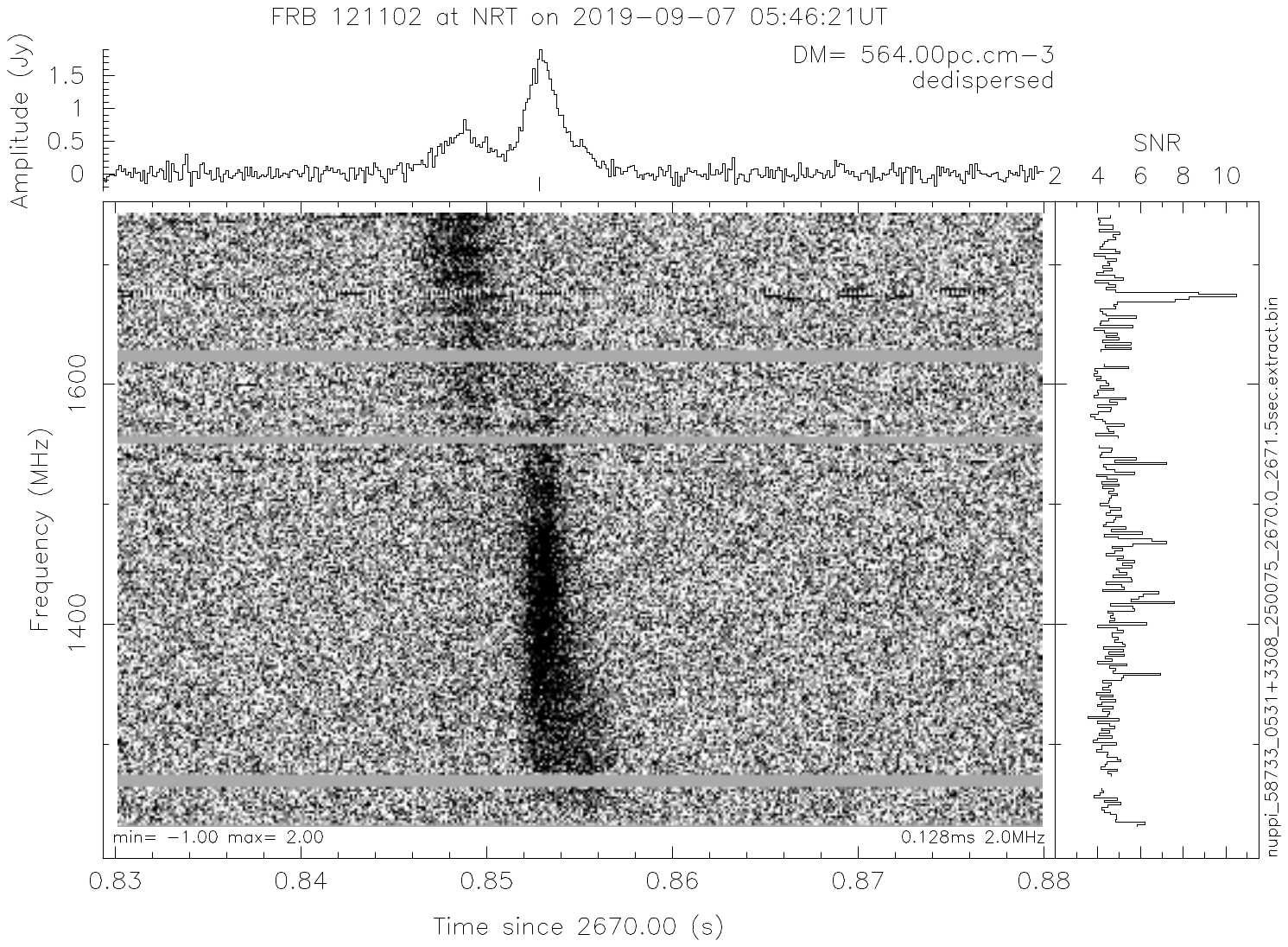}
      \caption{Waterfall (frequency-time) plot of one of the brightest bursts detected at NRT on 2019-09-07. The lateral panel on the top shows the Amplitude in Jy as a function of time, while the lateral panel on the right shows the spectrum. }
         \label{fig:NRT-waterfall}
\end{figure}

%Emitting band, morphology
Some of the NRT-detected bursts emit in the full NRT band (1230--1740\,MHz), while others are confined in specific ranges of this window. Table~\ref{tab:NRTbursts} list the emitting range for each of the NRT-detected burst. On average, the NRT bursts have a median emission bandwidth of 360\,MHz. 
This is in slight contrast to the Arecibo data set used in \citet{Aggarwal2021} which showed a lack of emission below 1300\,MHz and a narrower median width of 230\,MHz.
Five NRT bursts (14\%) show multi-component or downward drifting morphology (see an example in Fig.~\ref{fig:NRT-waterfall}), while the majority of the NRT bursts appear to be Gaussian-like bursts. We do not observe complex bifurcating structures as seen in the MeerKAT data set used in \citet{Platts2021}. 
%Although our ability to properly study burst morphology is likely limited by the time resolution of the downsampled NRT data which is typically 0.256\,ms \textcolor{red}{(Why are there different values??)}

%Figure~\ref{fig:BW-Fluence} is a plot of the observed emitting bandwidth of the NRT bursts vs the fluence, which shows a possible positive correlation.  

%\textcolor{red}{I am not sure what this really means. Are we saying the bandwidth of a burst is not necessarily limited, but possibly just a detection bias? As in there could be some variations in the spectral indices, but we can't be sure that the bursts are definitely not emitting in those bands??} 
%\begin{figure}
%   \centering
%   \includegraphics[width=\hsize]{BW-Amp.png}
%      \caption{Emitting bandwidth vs peak amplitude of the NRT bursts. \textcolor{red}{Try plotting fluence instead of Amp. Waiting for the data.}}
%         \label{fig:BW-Fluence}
%\end{figure}

\section{Discussion} \label{sec:discussion}
FRB~20121102 has been found to be in a star forming region \citep{Bassa2017} of a host galaxy of low metallicity \citep{Tendulkar2017}. These clues are consistent with the hypothesis of a magnetar progenitor model for FRB~20121102, since these environments also favour sources like long gamma-ray bursts and hydrogen-poor SLSNe-I which are considered to be associated with magnetar births.
%Theory
Theoretically, models of FRBs from magnetars predict small energy ratio between radio and high energy emission, as the radio energy budget is considered a small fraction of the total emitted energy. 
For example, \citet{Lyutikov2002} estimated a radio-to-X-ray energy ratio of $\sim 10^{-4}$ based on analogies to solar flares. \citet{margalit2020} expected a similar radio-to-gamma-ray ratio, while \citet{Lyubarsky2014} suggested the energy ratio could be at 10$^{-5}$--10$^{-6}$ based on  synchrotron maser interaction between the magnetic shock and the wind nebula of a magnetar.
%SGR1935 20-200 keV
In our Milky Way, magnetars are known to produce both X-ray and Gamma-ray bursts. 
A radio-to-X-ray energy ratio of $\sim2\times10^{-5}$ was reported for the simultaneous radio and X-ray (20--200\,keV) detection from the FRB-like burst of magnetar SGR~1935+2154 \citep{Mereghetti2020}. The spectrum of this burst was harder than those of typical Soft $\gamma$-ray repeater (SGR). These are motivations that magnetar-related gamma-ray detection of FRBs could be achieved.

%In addition, the case of magnetar SGR~1935+2154 which emitted an FRB-like radio signal at the same time as high energy bursts (radio-to-gamma-ray fluence ratio of $\eta$=9$\times$10$^{11}$\,Jy\,ms\,erg$^{-1}$\,cm$^{2}$ in the range of 20--200\,keV, see \citet{Mereghetti2020}) is a strong motivation that magnetar-related gamma-ray detection of FRBs could be achieved. 

\begin{figure}
   \centering
   \includegraphics[angle=0,width=1\hsize]{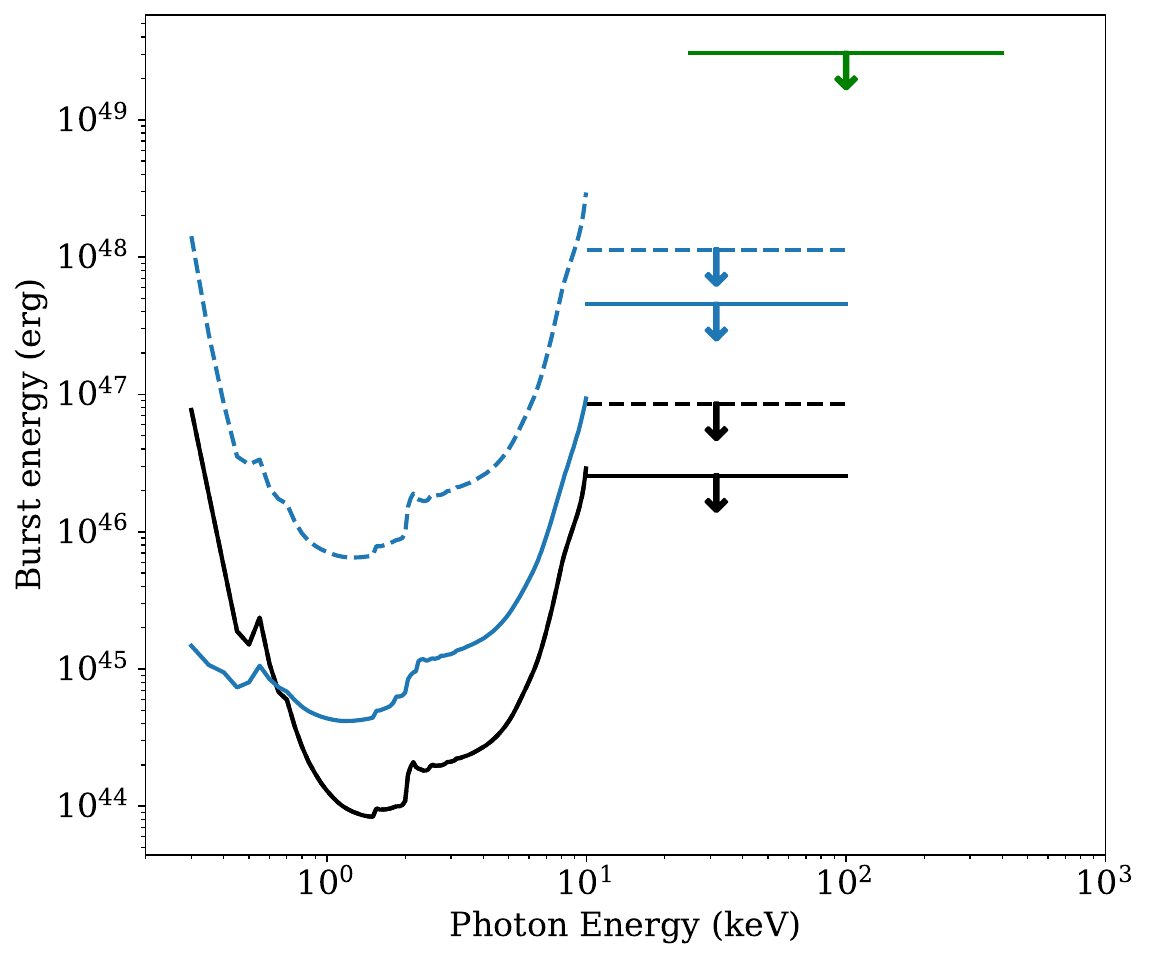}
      \caption{Limits on the energy of X-ray and gamma-ray bursts at the time of radio bursts from FRB~20180916 (in black; from \citet{Scholz2020}) and FRB~20121102 (in blue; from \citet{Scholz2017} and in green; this work). The 0.5--10\,keV data are from Chandra, 10--100\,keV from Fermi/GBM and 25--400\,keV from \textit{INTEGRAL}. The dashes lines indicate 5-$\sigma$ upper limits for single radio burst whereas the solid lines indicate stacking the multiple detected radio bursts.}
         \label{fig:UpperLimit}
\end{figure}

%XMM and Chandra
In the literature and for FRB~20121102 specifically, \citet{Scholz2017} conducted simultaneous radio and X-ray observations which led to a 5-$\sigma$ fluence upper limit of 3$\times$10$^{-11}$\,erg\,cm$^{-2}$ in the 0.5–10\,keV Chandra band for burst durations $<$700\,ms and a 5-$\sigma$ upper limit of  4$\times 10^{-9}$\,erg\,cm$^{-2}$ in 10–100\,keV using Fermi. 
Assuming isotropic emission, these limits correspond to burst energy of the order of 10$^{45}-10^{47}$\,erg (see blue lines in Fig.~\ref{fig:UpperLimit}) at the distance of FRB~20121102, which is measured to be at redshift $z$ = 0.193 which equates to a luminosity distance of 972\,Mpc \citep{Tendulkar2017}. 
They derived a radio-to-X-ray energy ratio of $ > 10^{-6}-10^{-8}$, depending on the spectral model and the absorbing column. 
Following the radio-to-gamma-ray fluence ratio formula $\eta = F_{\mathrm{1.4\,GHz}}/F_{\gamma}$ as defined in \citet{Tendulkar2017}, \citet{Scholz2017} 
place a lower limit of $\eta>6\times10^{9}$\,Jy\,ms\,erg$^{-1}$\,cm$^{2}$ for FRB~20121102, although this is subjected to the extrapolation from soft X-ray to gamma-ray and included the assumption that the radio fluence is the same at 1.4\,GHz and 2\,GHz. If compare to the Fermi measurements at 10–100\,keV, they obtain $\eta>2\times10^{8}$\,Jy\,ms\,erg$^{-1}$\,cm$^{2}$.
%---> and what does that imply?
%DeLaunay2016 -- 15-150 keV
\citet{DeLaunay2016} found two bursts from FRB~20121102 that occurred within the field of view of Swift BAT which resulted in gamma-ray upper limit of the order of 2$\times$10$^{-6}$\,erg\,cm$^{-2}$ and $\eta>10^{4.6}$\,Jy\,ms\,erg$^{-1}$\,cm$^{2}$ in the range of 15–150\,keV.

%This work (25-400 keV) radio-gamma
In this work, we present 5-$\sigma$ upper limit of $2.7\times 10^{-7}\rm{erg\, cm^{-2}}$ in the 25--400\,keV energy range for contemporary radio and gamma-ray bursts, which is a more stringent limit than \citet{DeLaunay2016} while being in comparable wavelengths. 
Considering the distance of FRB~20121102, this translates to a gamma-ray burst energy upper limit of 3$\times$10$^{49}$\,erg (see green line in Fig.~\ref{fig:UpperLimit}).
Assuming the same radio fluence of $F_{\mathrm{1.4\,GHz}}$=1.2\,Jy\,ms as reported in \citet{Tendulkar2017}, we derive $\eta>4\times10^{6}$\,Jy\,ms\,erg$^{-1}$\,cm$^{2}$ from our \textit{INTEGRAL} campaign.
This is still compatible with the radio-to-gamma-ray fluence ratio of a source similar to the Galactic magnetar SGR~1806$-$20, with $\eta_{\mathrm{SGR}}<10^{7}$\,Jy\,ms\,erg$^{-1}$\,cm$^{2}$ \citep{Tendulkar2016}.
On the other hand, our limit excludes the vast majority ($>$95\%) of the short ($T_{90}$<2\,s), the long (2\,s<$T_{90}$<100\,s) and the very long GRBs ($T_{90}$<100\,s) based on the fluence distribution of GRBs published in \citet{Bhat2016}.
This rules out the simultaneous presence of GRB and the radio bursts detected from the 2019 campaign of FRB~20121102, although we cannot exclude GRB being a progenitor of FRB~20121102 \citep{Margalit2019}.
Our results also do not support an event like the claimed detection of FRB~20131104 ($\eta=6\times10^{5}$\,Jy\,ms\,erg$^{-1}$\,cm$^{2}$) reported by \citet{DeLaunay2016}.

%Fermi 100MeV-300GeV - but no radio detection so drop it.
%Using the timestamps of the 235 bursts from FRB~121102 collected by \citet{Rajwade2020}, \citet{Principe2023} obtained a 2-$\sigma$ (95\%) upper limit of 2.72$\times$10$^{44}$\,erg\,s$^{-1}$ in the range of 100\,MeV and 300\,GeV using Fermi data.

%We also obtain a 5-$\sigma$ upper limit of $3.8\times 10^{-11}\rm{erg\, cm^{-2}}$ for permanent emission in the 25--100\,keV energy range. 
%Our upper limit.. Not very stringent, but the first contemporary limit at the 20–100\,keV range. Does not rule out the models mentioned above. 

The gamma-ray non-detection could be due to the extragalactic distance of the FRB, resulting in gamma ray emission below the detectability of \textit{INTEGRAL}. In addition, different beaming angles between the radio and high emission has been proposed by \citet{Sridhar2021}, meaning these multi-wavelength signals might not be correlated. It is also possible that the gamma-ray emission is heavily attenuated by the local environment, which could be the case as FRB~20121102 co-locates with a persistent radio source \citep[PRS;][]{Chatterjee2017} that might be a result of a pulsar wind nebula (PWN) driven by the young magnetar \citep{Kashiyama2017}.
Indeed, \citet{Chen2023} conducted a comprehensive study of the PRS and found that the inferred size, and the flat radio spectrum favors a plerion origin.
Alternatively, \citet{Metzger2017} proposed a scenario explained by the interaction between the supernova blast wave and the surrounding progenitor wind nebula.
Finally, it is also possible that the radio bursts are amplified due to lensing events by the intervening medium \citep{Cordes2017}, meaning that the high energy emission could be much lower if the radio-to-high-energy fluence ratio holds. 

The recently launched gamma-ray instrument known as the 
 space-based multi-band astronomical variable objects monitor \citep[SVOM;][]{SVOM} could provide improved sensitivity in the 15\,keV--5\,MeV range which would lead to more stringent radio-to-gamma-ray fluence limit. We encourage further multi-wavelength campaigns on FRBs, particularly the nearby, low redshifts repeating sources.

%Optical 
We also note that an optical burst fluence upper limit of 46\,mJy\,ms was reported by \citet{Hardy2017} for FRB~20121102 during a period of time when 13 radio bursts were detected. Our optical limit of 12\,mJy\,ms, although more strigent, was taken during a radio quiet period where no radio burst was detected. Hence we do not have any knowledge on simultaneous radio/optical bursts and cannot compare our results directly with those in the literature.

\section{Conclusions} \label{sec:conclusion}
We did not detect any contemporary bursts from FRB~20121102 in hard X-ray/soft $\gamma$-ray at the times when radio bursts were detected by the NRT and Arecibo during observations taken in 2019.
Using the \textit{INTEGRAL} satellite, we obtain a 5-$\sigma$ upper limit of $2.7\times 10^{-7}\rm{erg\, cm^{-2}}$ in the 25--400\,keV energy range for contemporary radio and high energy bursts, and a 5-$\sigma$ upper limit of $3.8\times 10^{-11}\rm{erg\, cm^{-2}}$ for permanent emission in the 25--100\,keV energy range.  The contemporary limit translates
to a gamma-ray burst energy upper limit of 3$\times10^{49}$\,erg and a fluence ratio of  $\eta>4\times10^{6}$\,Jy\,ms\,erg$^{-1}$\,cm$^{2}$.
Optical observations taken at the OHP resulted in an upper limit of $R$=22.2\,mag and a flux limit of 12\,mJy\,ms, although these were taken during a radio quiet period in 2017.
In addition, we report 131 radio bursts detected by the NRT between 2017--2021, which provides a more precise measurement on the activity periodic window of 154$\pm{2}$ days.

\begin{acknowledgements}
This manuscript is written in memory of Dr. Christian Gouiffes, who led this multi-wavelength observation campaign of FRB~20121102. Dr. Gouiffes unexpectedly passed away in 2023. 
The Nançay Radio Observatory is operated by the Paris Observatory, associated with the French Centre National de la Recherche Scientifique (CNRS). 
OHP is a department of Pytheas Institute in Marseille, co-directed by CNRS and Aix-Marseille University. 
This paper is based on observations with INTEGRAL, an ESA project with instruments and science data center funded by ESA member states (especially the PI countries: Denmark, France, Germany, Italy, Switzerland, Spain) and with the participation of Russia and the USA. E.~O'Connor is thanked for his help with the GASP observations.
We are very grateful for the help from Paul Scholz for producing this updated version of Figure~\ref{fig:UpperLimit}.
\end{acknowledgements}

\bibliographystyle{aa}
\bibliography{ref}

\end{document}